\documentstyle[psfig]{mn}
\newif\ifAMStwofonts
\AMStwofontstrue

\def\mks{{\rm\umu s}}

\ifoldfss
  \ifCUPmtlplainloaded \else
    \NewTextAlphabet{textbfit} {cmbxti10} {}
    \NewTextAlphabet{textbfss} {cmssbx10} {}
    \NewMathAlphabet{mathbfit} {cmbxti10} {} 
    \NewMathAlphabet{mathbfss} {cmssbx10} {} 
  \fi
  \ifAMStwofonts
    \ifCUPmtlplainloaded \else
      \NewSymbolFont{upmath} {eurm10}
      \NewSymbolFont{AMSa} {msam10}
      \NewMathSymbol{\upi}     {0}{upmath}{19}
      \NewMathSymbol{\umu}     {0}{upmath}{16}
      \NewMathSymbol{\upartial}{0}{upmath}{40}
      \NewMathSymbol{\leqslant}{3}{AMSa}{36}
      \NewMathSymbol{\geqslant}{3}{AMSa}{3E}

       \let\le=\leqslant
       \let\ge=\geqslant
    \fi
  \fi
\fi 

\ifnfssone
  \newmathalphabet{\mathit}
  \addtoversion{normal}{\mathit}{cmr}{m}{it}
  \addtoversion{bold}{\mathit}{cmr}{bx}{it}
  \newmathalphabet{\mathbfit} 
  \addtoversion{normal}{\mathbfit}{cmr}{bx}{it}
  \addtoversion{bold}{\mathbfit}{cmr}{bx}{it}
  \newmathalphabet{\mathbfss} 
  \addtoversion{normal}{\mathbfss}{cmss}{bx}{n}
  \addtoversion{bold}{\mathbfss}{cmss}{bx}{n}
  \ifAMStwofonts
    \ifCUPmtlplainloaded \else
      \UseAMStwoboldmath
      \makeatletter
      \new@mathgroup\upmath@group
      \define@mathgroup\mv@normal\upmath@group{eur}{m}{n}
      \define@mathgroup\mv@bold\upmath@group{eur}{b}{n}
      \edef\UPM{\hexnumber\upmath@group}
      \new@mathgroup\amsa@group
      \define@mathgroup\mv@normal\amsa@group{msa}{m}{n}
      \define@mathgroup\mv@bold\amsa@group{msa}{m}{n}
      \edef\AMSa{\hexnumber\amsa@group}
      \makeatother
      \mathchardef\upi="0\UPM19
      \mathchardef\umu="0\UPM16
      \mathchardef\upartial="0\UPM40
      \mathchardef\leqslant="3\AMSa36
      \mathchardef\geqslant="3\AMSa3E

       \let\le=\leqslant
       \let\ge=\geqslant
    \fi
  \fi
\fi 

\ifnfsstwo
  \DeclareMathAlphabet{\mathbfit}{OT1}{cmr}{bx}{it}
  \SetMathAlphabet\mathbfit{bold}{OT1}{cmr}{bx}{it}
  \DeclareMathAlphabet{\mathbfss}{OT1}{cmss}{bx}{n}
  \SetMathAlphabet\mathbfss{bold}{OT1}{cmss}{bx}{n}
  \ifAMStwofonts
    \ifCUPmtlplainloaded \else
      \DeclareSymbolFont{UPM}{U}{eur}{m}{n}
      \SetSymbolFont{UPM}{bold}{U}{eur}{b}{n}
      \DeclareSymbolFont{AMSa}{U}{msa}{m}{n}
      \DeclareMathSymbol{\upi}{0}{UPM}{"19}
      \DeclareMathSymbol{\umu}{0}{UPM}{"16}
      \DeclareMathSymbol{\upartial}{0}{UPM}{"40}
      \DeclareMathSymbol{\leqslant}{3}{AMSa}{"36}
      \DeclareMathSymbol{\geqslant}{3}{AMSa}{"3E}

       \let\le=\leqslant
       \let\ge=\geqslant
    \fi
  \fi
\fi 

\ifCUPmtlplainloaded \else
  \ifAMStwofonts \else 
    \def\upi{\pi}
    \def\umu{\umu}
    \def\upartial{\partial}
  \fi
\fi

\title{Further daemon detection experiments}
\author[E. M. Drobyshevski]
       {E. M. Drobyshevski\\
        A.F.Ioffe Physical-Technical Institute, Russian
        Academy of Sciences, 194021 St Petersburg, Russia}
\date{Accepted ---.
      Received ---;
      in original form ---}

\pagerange{\pageref{firstpage}--\pageref{lastpage}}
\pubyear{2000}

\begin{document}

\maketitle

\label{firstpage}

\begin{abstract}
The experiments on detection of daemons captured into geocentric
orbits, which are based on the postulated fast decay of
daemon-containing nuclei, have been continued. By properly
varying the experimental parameters, it has become possible to
reveal and formulate some relations governing the interaction of
daemons with matter. Among them are, for instance, the emission
of energetic Auger-type electrons in the capture of an atomic
nucleus, the possibility of charge exchange involving the capture
of a heavier nucleus etc. The decay time of a daemon-containing
proton has been measured to be
$\Delta\tau_{\rm ex} \approx 2\,\mks$. The daemon flux at
the Earth's surface is
$f_\oplus \sim 10^{-7}\,{\rm cm^{-2}s^{-1}}$. One should point
out, on the one hand, the reproducibility of the main results,
and on the other, the desirability of building up larger
statistics and employing more sophisticated experimental methods
to reveal finer details in the daemon interaction with matter.
\end{abstract}


\section{Introduction. General ideology and the results of the
previous experiment}

We reported earlier (Drobyshevski 2000c) on experimentally
detecting indications of the existence of a flux of
strongly-penetrating, nuclear-active particles moving with a
velocity $\la\!10$ km/s. The experiment supported our
expectations aimed at a search for hypothetical DArk Electric
Matter Objects, daemons, i.e. elementary Planckian black holes,
which carry several (up to $Z_{\rm d}\approx 10$) electronic
charges and are part of the DM of the Galactic disk and,
possibly, of the Universe as a whole.

Because of their large mass ($\approx\!2\cdot 10^{-5}$ g), daemons
possess a giant penetrating ability. Nevertheless, as a result of
the decelerating action of the matter they build up inside
celestial bodies of the type of the Sun and the planets, as
well as in helio- and planetocentric orbits. Due to their large
charge, negative daemons are capable of catalyzing the fusion of
light nuclei. The daemon-assisted proton fusion supplies
apparently most of the Solar energetics and accounts for the
electron-capture neutrino deficiency in the Solar flux
(Drobyshevski 1996b). Bearing in mind the relativistic nature of
the daemon, one could expect that the assumption of possible
poisoning of its catalytic properties in the capture of heavy
nuclei, similar to that expressed with respect to other particles
(quarks etc.) (see, e.g., Salpeter 1970), would not prove to be
correct. It could be anticipated that such nuclei (or their
components) would either escape under the relativistic horizon of
the daemon (Drobyshevski 1996a) or be ejected as a result of the
decay of the component nucleons into which the daemon penetrated
(Drobyshevski 2000a,b).

As follows from a comparison of the number of daemons captured by
the Sun from the Galactic disc with their number required for the
catalysis necessary to support the Solar luminosity, the time
needed for a daemon to recover its catalytic properties inside
the Sun is about $10^{-7}-10^{-6}$ s (Drobyshevski 2000a,b). It
remains unclear what is the real mechanism by which the daemon
frees itself of a heavy nucleus. One could expect that the
situation would become clearer as the experiments are continued
in the direction chosen by us.

We assembled (Drobyshevski 2000c) an extremely simple setup for
daemon detection based on the above two mechanisms of interaction
of negative daemons with light ($Z_{\rm n} \le Z_{\rm d} \approx
10$) and heavy nuclei. It consists essentially of two transparent
polystyrene plates, 4 mm thick and $0.5\times 0.5\,{\rm m^2}$ in
area, coated on the bottom by a relatively thin ($3.0-3.5\,{\rm
mg/cm^2}$) layer of the ZnS(Ag) phosphor (its Ag content is
$\sim\!0.5\cdot 10^{-4}$ g atom/mole ZnS). For an average
phosphor grain size $\sim\!12\,{\rm\umu m}$, this density
provides a close-to-monolayer coverage with certain gaps between
the ZnS(Ag) grains. The plates are light-isolated from one
another by black paper and are mounted with a 7-cm spacing at the
center of a cubic box of tinned sheet with an edge of 52 cm. Each
plate is viewed by a PM tube from above and from below. Four
identical modules, $1\,{\rm m^2}$ in total area and arranged side
by side in the same horizontal plane, were built altogether. We
readily see that the system is asymmetric with respect to the
vertical direction, thus permitting one to judge of the daemon
flux direction and to draw conclusions on some properties of the
radiations measured.

In the absence of a flux of particles causing scintillations,
there should be no correlation between the signals of the PM
tubes viewing the top and the bottom plates. In these conditions,
the distribution $N_2(\Delta t)$ of the fairly rare (noise etc.)
signals of the second (bottom) PM tube in the time shift $\Delta
t$ between their beginning and the beginning of the signal from
the first (top) PM tube can be approximated with a constant. An
exception to this are the cosmic rays producing a strong maximum
at $\Delta t = 0$, with the signals appearing frequently
simultaneously in several modules. What we found, however, was a
statistically significant maximum at $\Delta t \approx 20-40\,\mks$,
which, while being totally unexpected from the
standpoint of standard cosmic-ray physics, is exactly what had
motivated the present experiment. It is due to signals from the
bottom PM tubes, which arrive after the signals from the top
(leading) PM tubes triggering the measurement system. One can
infer also an existence of a leading maximum, i.e. a maximum
with $\Delta t < 0$. It is much less significant and is shifted
by $\Delta t \approx -(40-60)\,\mks$. It is such maxima that
were anticipated by us based on the daemon hypothesis.

These observations, besides indicating the existence of a new
type of penetrating cosmic radiation, suggested that the velocity
of the discovered particles is barely $\sim\!5-10$ km/s, which
should be characteristic of objects captured into geocentric
orbits with a perigee inside the Earth. Their flux through the
Earth's surface reaches $\sim\!10^{-8}-10^{-7}\,
{\rm cm^{-2}s^{-1}}$.

We are presenting below a description of further experiments,
which support our preliminary conclusions of the existence of
daemons and shed light on some features of their interaction with
matter.

However prior to going over to these results, we shall formulate
the main postulates which permit a noncontradictory
interpretation of our experimental data within the daemon
concept. These postulates have formed to a considerable extent in
the course of an analysis of the results themselves made on the
basis of fairly general physical ideas.

\section{Main assumptions concerning the interaction of negative
daemons with matter}

\noindent (a) A slowly-moving ($V \la 100$ km/s) daemon
impact neither ionizes nor excites the electronic shells of
atoms. Therefore it cannot produce scintillations.

\noindent (b) In passing through matter, a daemon can capture an
atomic nucleus. The ensuing drop of the nucleus to deeper levels
in the electric field of the daemon will cause an Auger-type
emission of electrons which initially surrounded the atomic
nucleus.

\noindent (c) As the daemon continues to propagate through
matter, it can capture a heavier nucleus while loosing the old
one (or its remnant, see below), again with emission of Auger
electrons. The capture of a heavier nucleus is similar to the
effect of charge exchange of ions moving through a gas, which is
well known from the physics of gas discharge. Therefore as a
daemon enters ZnS(Ag) from air or polystyrene, one should expect
the light nuclei to become replaced by the S, Zn, and Ag nuclei
and excitation of a scintillation. In this case the process of
`charge exchange' is more efficient for a larger mass
difference between the exchanging nuclei.

\noindent (d) If the nucleus captured earlier by the daemon has a
small mass and $Z_{\rm n} \le Z_{\rm d}$, interaction of this
complex with another light nucleus and the capture of the latter
may culminate in nuclear fusion. Because of the internal
conversion, the fusion energy has a high probability of becoming
converted to the kinetic energy of the compound-nucleus thus
formed, which escapes from the daemon. The daemon remains free
afterwards for some time.

\noindent (e) If the daemon captures a light nucleus in a
sufficiently rarefied (noncondensed) medium, for instance, in
air, the light nucleus will not be able to shed the excess energy
and, thus, may turn out to occupy such a high level as not even
to be in contact with the daemon. (Recall that the lowest
nucleus-daemon level for $Z_{\rm d} = 10$ lies within the
nucleus with $A \ge 2$, and inside the nucleon for $Z_{\rm n} \ge
24/Z_{\rm d}$.)

\noindent (f) It appears that the scintillations observed by us
are primarily caused by the Auger electrons emitted in the
capture of new atomic nuclei. We cannot at present maintain with
certainty that what we detect are the nuclei released by the
daemon and representing products of the fusion of light nuclei or
particles produced in the decay of daemon-containing nucleons (or
nuclei), although such events must certainly take place.
Therefore we believe that as a daemon passes through a
scintillator layer, the scintillation is primarily excited in the
capture of a nucleus from the latter by the daemon, and by the
resultant emission of Auger electrons. Scintillations are also
produced in Auger electron emission and in the capture of a
nucleus from outside the scintillator, if part of these
electrons reaches the scintillator.

\noindent (g) The daemon-containing heavy nucleus decays
apparently not in an explosive manner as we believed before
(Drobyshevski 2000a,b), but rather step by step, nucleon by
nucleon. Because a negative daemon should preferably reside in
the protons of a nucleus, the decay most probably occurs in the
proton-by-proton manner. The proton-by-proton disintegration
should be accompanied by emission of excess neutrons from the
nucleus. This gradual `digestion' of the nucleus ends probably
either in the decay (absorption?) of the last proton of the
remaining tritium (or even of $^4{\rm H}$) or in a capture of an
encountered nucleus which is heavier than the disintegrating
residual nucleus. Clearly enough, the nuclear decay time and the
recovery by the daemon of its postulated catalytic properties
are, on the whole, proportional to the initial nuclear mass;
however in the presence of background matter from which another
nucleus can be captured (for digestion or fusion with the
remainder of the old one) this time depends to some extent on the
nuclear properties of the surrounding matter.

While the first postulates are based on the relations known from
atomic and nuclear physics, the latter (g) postulate follows
already from our experiments.

\section{Increasing the experiment duration}

The discovery of a nuclear-active radiation propagating with a
velocity on the km-scale, i.e., of a population captured into
particularly geocentric orbits, was somewhat unexpected for us.
Therefore the next natural step was to increase the observation
time of the signals from the bottom scintillator surface from
$-100 \le \Delta t \le +100\,\mks$ (Drobyshevski 2000c) to
$-250 \le \Delta t \le +250\,\mks$ with respect to the
signal from the top scintillator. We hoped that beyond the $\pm
100\,\mks$ interval we would enter the zone of purely random
signals with a $N_2(\Delta t) = {\rm const}$ distribution.
However this conjecture did not gain a support.

We start our consideration with $\Delta t < 0$, i.e., with
signals belonging to an upward flux of non-monoenergetic (see
Fig.~1) particles. Here, except a taking shape maximum at
$-40 \le \Delta \le -20\,\mks$, which corresponds to the 7
cm interscintillator distance at the particle normal velocity of
$\sim\!2-3$ km/s, one clearly sees another three maxima, with the
first one, fairly diffuse, extending from -220 to -160 $\mks$,
the second from -120 to -100 $\mks$, and the third, from -80 to
-60 $\mks$. Bearing in mind our assumptions (see Sec.2), the
first maximum can be due to the bottom scintillator excited by
the Auger electrons, which are emitted in the capture by daemons
of air nuclei after getting freed of the Fe and Sn heavy nuclei
captured in the traversal of the lower part of the tinned-sheet
casing at various angles. The Auger electrons can be emitted also
in the capture of Sn from the thin tin coating of the iron sheet.
After the participation in all these processes, part of the
daemons reach the top luminophor, and it is here that they
generate the trigger pulse correlating with the signals of the
first diffuse maximum.

\begin{figure}
\psfig{file=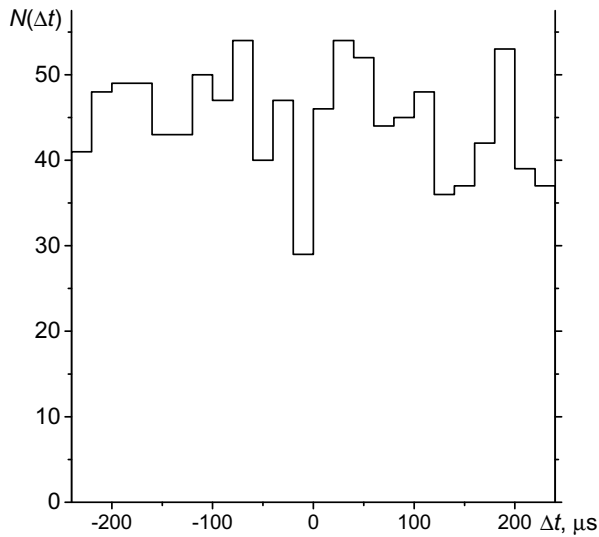,width=8cm}
\caption{Distribution of the lower scintillator plate signals vs
their time shift relative the top scintillator signals. Time of
exposition is $\approx$1000 hours.}
\end{figure}

The sharper second and third maxima are possibly produced as the
daemon captures Ag ($Z_{\rm n} = 47$) and Zn ($Z_{\rm n} = 30$)
nuclei, respectively, in traversing the bottom luminophor. Note
that the captured S nucleus has a high probability to be shortly
lost when it encounters a Zn nucleus, and the probability of
capturing Ag is not small despite the low concentration of Ag,
because for a velocity of $\sim$2 km/s the daemon free path
until meeting an Ag atom is comparable with the ZnS(Ag) grain
size. On traversing at any angle the bottom polystyrene plate
into the space before the top luminophor, daemons finish
digesting the Ag and Zn nuclei, so that when they capture air
atoms, they emit here Auger electrons, which trigger the top
luminophor. If our scenarium is true, the time shift between
these maxima relative to $\Delta t = 0$ provides an idea about
the time per decay of a daemon-containing proton. It turns out to
be $\Delta\tau_{\rm ex} \approx \tau_{\rm ex} / n_{\rm p} =
\tau_{\rm ex}/Z_{\rm n} \approx 2\,\mks$, which shows that our
estimates of $\tau_{\rm ex}$ based on Solar energetics
(Drobyshevski 2000a,b) were clearly underestimates.

The position of the first maximum ($\Delta t \approx 180-200\,
\mks$) yields $\sim$1.5 km/s for the velocity with which a
daemon passes normally the distance of 29 cm from the lower box
cover to the upper luminophor layer. Taking into account the
random angular distribution of the trajectories, the average
velocity increases to $\sim$3 km/s.

For $\Delta t > 0$, the daemons move downward.

The first distinct maximum at $\sim\!20 < \Delta t < 40 \mks$
corresponds apparently to the time needed for a daemon to cross
the 7-cm gap separating the scintillator layers. For an arbitrary
trajectory inclination, this time, as we saw before (Drobyshevski
2000c), corresponds to a velocity $\sim\!3-5$ km/s, i.e., to
objects in geocentric orbits with a perigee inside the Earth. The
interpretation of this maximum, as well as the $-40<\Delta
t<-20\,\mks$ maximum, should be approached with a certain caution,
because the fairly slow digestion of nuclei by a daemon could
give rise to a certain bias toward inclined trajectories. A
daemon poisoned by a poorly digested heavy nucleus is less likely
to capture a Zn or Ag nucleus from the next scintillator layer
and to initiate a scintillation in it if it propagates by the
fastest (direct) path. Another selection arises because of the
higher probability of capture of Ag, despite its low content, by
a slower-moving daemon (the capture cross section is
proportional to $V^{-2}$).

Because of the ZnS(Ag) coating being not continuous, daemons with
nuclei captured both in the upper scintillator have a fairly high
probability of turning up also in the space under the bottom
scintillator. Therefore in order to reliably separate and
subsequently identify the maxima according to their origin (the
end of digestion of a Zn or Ag nucleus captured in the top or
bottom scintillator etc.) for $\Delta t \ge 40-60\,\mks$, the statistics of
the events should be much larger than they are presently.
Nevertheless, there are grounds to believe that the maximum which
is just barely discernible near +100 ms is due to the decay of
the Ag captured in the top ZnS(Ag) scintillator, and that the
maximum at $\sim\!180 < \Delta t < 200\,\mks$ for the daemons
propagating downward, has the same origin as the one at $-160>
\Delta t > -180\,\mks$, i.e., interaction of the upward-moving
daemons with the tin coating of the lower part of the box (in
this case, however, the Auger electrons are certainly emitted
from the thin tin layer in the capture of Sn nuclei). There is no
Zn maximum in the $+(40-80)\,\mks$ interval, because most of the
captured Zn nuclei disintegrate before entering the second
scintillator layer. Recall that we selected in the oscillograms
only the events that contained one signal.

\section{Experiment with Additional Tinned-Iron Sheets}

To verify once more that what we detect in these experiments are
not artifacts but real reproducible events, we placed into the
gap between the top and bottom phosphor layers a 0.25-mm thick
sheet of tinned iron in all the four modules. The thickness of
the tin layer on each side of the sheet was $\sim\!2\,{\rm\umu m}$.
This sheet was placed directly on the black paper providing
optical isolation of the scintillator layers from one another, so
that it was 5 mm above the bottom scintillator.

The experiment was run for 550 h. It was performed after the
experiment with the $\pm 100\,\mks$ time interval (Drobyshevski
2000c) and before the one with the $\pm 250\,\mks$ interval, which
is described in Section 3 and demonstrates also the
reproducibility of the main results within the $\pm 100\,\mks$
interval.

We expected that the tinned sheet, because of the presence in it
of Fe and Sn nuclei, would strongly affect the $N_2(\Delta t)$
distribution obtained in the conditions where heavy nuclei are
contained only in the ZnS(Ag) layers.

This was exactly what happened. The distribution for $\Delta t <
0$ became substantially more monotonic (Fig.~2). It no longer
exhibits individual, fairly sharp maxima evident in Fig.~1. One
sees actually one, closely filled maximum extending from -160 to
0 $\mks$. If the statistics were larger, there would probably
appear traces of the old maxima. As it is, one can only maintain
that charge exchange in the thin tin layer of the additional
tinned sheet accompanied by a loss of (remnants of) Zn and Ag
nuclei and a capture of $\sim\!^{119}{\rm Sn}$ heavy nuclei
proceeds apparently so efficiently that the upward moving daemons
deliver without loss only remnants of these Sn nuclei to the top
scintillator. And it is only here that these remnants are
replaced by Ag (and Zn) nuclei to initiate the triggering
scintillation. Recalling that the distribution for $\Delta t<0$
is filled up to $\Delta t \rightarrow 0$, one could conjecture
that some (triggering) scintillations in the top ZnS(Ag) layer
are caused by the Auger electrons escaping from the thin top tin
layer of the sheet when a daemon captures a Sn nucleus in it. We
may recall that electrons of energy $\sim\!0.1-1$ MeV are capable
of passing through a 2-${\rm\umu m}$ layer of tin while being
stopped by a 0.25-mm layer of iron or by 4 mm of polystyrene.

\begin{figure}
\psfig{file=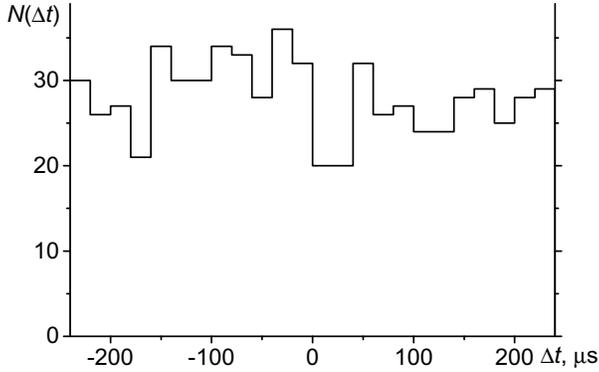,width=8cm}
\caption{The same as figure 1, but the tinned-iron sheet was laid
on the lower scintillator plate.}
\end{figure}

The part of the $N_2(\Delta t)$ distribution for $\Delta t > 0$,
which characterizes the downward daemon flux also differs from
the corresponding distribution in Fig.~1.

Note the dip at $0 < \Delta t < 40\,\mks$, after which a peak at
$40 < \Delta t < 60\,\mks$ appears. This dip can be readily
interpreted as due to screening by the tinned sheet of the bottom
ZnS(Ag) scintillator. In passing through the sheet, the daemon
becomes poisoned by, say, a Fe nucleus ($Z_{\rm n} = 26$), and
therefore does not excite the ZnS(Ag) layer. Only after the
traversal of the ZnS(Ag) layer and digestion of the iron nucleus
in $\Delta t \approx 26\times 2\,\mks \approx 50\,\mks$ the daemon
captures a nucleus in the air and generates Auger electrons, that
excite the bottom scintillator layer. Therefore even in the
presence of fast daemons the width of the dip should not depend
on specific features of their velocity distribution.

The depth of the dip at $0 < \Delta t < 40\,\mks$ characterizes
apparently the noise level in the given $N_2(\Delta t)$
distribution. But then, as follows from the $N_2(\Delta t)$
histogram in Fig.~2, the number of events initiated by daemons
during the time of observation is 193 (taking into account
errors, it could be reduced to $\sim$one third this figure). This
yields for the measured daemon flux $f_\oplus \approx
(0.3-1)\cdot 10^{-8}\,{\rm cm^{-2}s^{-1}}$. In view of the fact
that the level ($U_1 \ga 2.5$ mV) of the triggering pulses is
three times that of the second-trace signals $U_2 \ge 0.8$ mV we
select, and that the frequency of the pulses is ~ proportional to
the cube of their amplitude, we obtain for a probable estimate
$f_\oplus\sim 10^{-7}\,{\rm cm^{-2}s^{-1}}$.

Accumulation of larger statistics will hopefully reveal in the
$N_2(\Delta t)$ dependence (for $\Delta t > 0$) maxima
corresponding to the decay of Sn, Ag, and Zn, as well as permit
determination of the daemon distribution in velocities from the
$N_2(\Delta t)$ plot obtained without iron shields.

\section{Some Conclusions and Prospects}

We have succeeded in developing a nonstandard method of detection
of daemons through the understanding of the fact that the
fundamental nonrelativistic nature of these particles permits
them to recover their catalytic properties in a relatively short
time by shedding the captured massive nucleus in a still unknown
manner. Using fairly simple, not to say primitive handy means, we
revealed definite and largely expected indications of the
existence of daemons. Their low velocity was not anticipated,
however. It is characteristic of a population captured by the
Earth. The flux of this population through the Earth's surface is
$f_\oplus\sim 10^{-7}\,{\rm cm^{-2}s^{-1}}$, possibly.

The results produced by varying the experimental conditions were
likewise largely anticipated, which suggests that our ideas
concerning the properties of daemons are, on the whole, correct,
and that we correctly understand what happens when we change the
parameters of the setup and the conditions of observations. A
similar experiment repeated months after the preceding one
demonstrates a good reproducibility.

A thorough analysis of the data obtained has permitted
formulation of a number of fairly obvious postulates governing
the behavior of daemons in matter (decay of nucleons and
recapture of heavier nuclei, Auger electron emission etc.).

Besides the confirmation of the discovery of daemons itself and
of their Earth-bound population, this work has culminated in a
fundamental result of an actually direct measurement of the decay
time of a daemon-containing proton. It was found to be
$\Delta\tau_{\rm ex} \approx 2\,\mks$. This value is larger by at
least an order of magnitude than the figure for the average time
for the daemon to shed a heavy nucleus ($\tau_{\rm ex} \sim
10^{-7}-10^{-6}$ s), which was obtained by us from an analysis of
the Solar energetics (we may recall that $\Delta\tau_{\rm ex}
\approx \tau_{\rm ex}/Z_{\rm n}$).

Because $\Delta\tau_{\rm ex} = {\rm const}$ characterizes the
decay time of one proton, we have actually discovered a
daemon-associated method of determining $Z_{\rm n}$ of the
substance through which it passes and whose nuclei it captures.

It should be admitted that because of the statistical nature of
the results we obtain we just have to work close to the
confidence level in order to reach as rich an understanding of
the essence of the observed phenomena as possible in a minimum
time. It is thus obvious that one has to accumulate as
statistically reliable material as possible while simultaneously
improving the detection system and varying its parameters. Richer
statistics will permit refining this important parameter,
$\Delta\tau_{\rm ex}$, as well as revealing and making more
reliable new features in the distribution of events happening
between remote scintillation detectors from the corresponding
time delays. Some of the conclusions made by us here will
possibly have to be modified. As a result, one will be able to
determine the daemon distributions in velocity and directions and
their long-period variations, as well as to reveal finer
details in the interaction of daemons with various components of
the material and, more generally, with matter as a whole. To
reach the latter goal, one would naturally have to use more
sophisticated techniques of nuclear physics than those employed
by us thus far. Obviously enough, daemons would themselves serve
as a completely new tool to probe processes on both the
subnuclear and cosmological scales.

\section*{Acknowledgments}

The author is greatly indebted to M.V.Beloborodyy, R.O.Kurakin,
V.G.Latypov and K.A.Pelepelin for assistance in experiments.

\label{lastpage}

\end{document}